# Digital Image Correlation used to Analyze the Multiaxial Behavior of Rubber-like Materials

by

L. Chevalier, S. Calloch, F. Hild and Y. Marco

LMT-Cachan

ENS de Cachan / CNRS – UMR 8535 / Université Paris 6

61 avenue du Président Wilson

94235 Cachan Cedex

France

# Digital Image Correlation used to Analyze the Multiaxial Behavior of Rubber-like Materials

by L. Chevalier, S. Calloch, F. Hild and Y. Marco

*Abstract*

We present an experimental approach to discriminate models describing the mechanical behavior of polymeric materials. A biaxial loading condition is obtained in a multiaxial testing machine. An evaluation of the displacement field obtained by digital image correlation allows us to evaluate the heterogeneous strain field observed during these tests. We focus on the particular case of hyper-elastic models to simulate the behavior of a rubber-like material. Different expressions of hyper-elastic potential are used to model experiments under uniaxial and biaxial loading conditions.

## 1. Introduction

The macromolecular structure of polymers leads to a specific behavior, in particular large strain levels are generally observed. A large variety of models exists to simulate the behavior of polymeric materials. These models give similar results for shear tests but lead to different responses in extension for example. For the simple case of rubber-like materials at room temperature, hyper-elasticity can be modeled in many different ways (Mooney 1940, Rivlin 1948, Gent and Thomas 1958, Hart Smith 1966, Alexander 1968). The material parameters are usually tuned in uniaxial tension but they give different responses for other loading conditions (i.e., shear stress or multiaxial loading).

Multiaxial testing allows one to discriminate different models (see for example Kawabata 1970, Meissner 1985). A specific testing machine called ASTREE enables us to apply a multiaxial loading along three perpendicular directions. Consequently, strain field heterogeneity may occur and does not lead to a simple identification. It follows that a displacement field measurement is needed to check and quantify the heterogeneity of the strain field. Optical measurement techniques are one of the most appealing methods to estimate two-dimensional displacement fields. Upon loading, the observed surface moves and deforms. By acquiring pictures for different load levels by using a Charge-Coupled Device (CCD) camera, it is possible to determine the in-plane displacement field by matching different zones of the two pictures. The simplest image-matching procedure is cross-correlation that can be performed either in the physical space (Peters and Ranson 1981, Sutton 1983, Chu et al. 1985, Mguil et al. 1998, Sutton et al. 2000) or in the Fourier space (Chen et al. 1993, Berthaud et al. 1996, Chiang et al. 1997, Collin et al. 1998). By using Fast Fourier Transforms (FFT), one can evaluate the cross-correlation function very quickly.

The multiaxial testing machine used in this study is presented in Section 2. Section 3 introduces the basics related to the correlation of two signals. The correlation algorithm (CORRELI$^{GD}$) is presented. We then discuss the choice of the strain tensor when the infinitesimal approximation is no longer valid. In the example of pure rotation, the precision of the correlation technique is assessed by using an exact solution within the framework of infinitesimal strains. Section 4 deals with an application using both ASTREE and CORRELI$^{GD}$ tools. In particular, the choice of the size of the zone of interest in the correlation technique is discussed when the displacements are not infinitesimal. An accurate identification of the hyper-elastic potential is performed in case of a rubber-like material: Smactane[TM].

## 2. A Multiaxial Testing Machine: ASTREE

The biaxial tests presented in this paper have been carried out on a triaxial testing machine ASTREE (Fig. 1). This electro-hydraulic testing machine has six servohydraulic actuators (Fig. 2). This machine has been developed by LMT-Cachan and Schenck AG, Darmstadt, Germany. The load frame is composed of a fixed base, four vertical columns and a mobile crosshead. The testing space 650 x 650 x 1500 mm$^3$ in volume, delimited by the six actuators, may be used for the tests. The two vertical actuators have a load capacity of 250 kN and a stroke range of 250 mm. The four horizontal actuators have a load capacity of 100 kN and a stroke range of 250 mm.

For efficient protection of the actuators at specimen failure (which can cause extremely high side and twist forces), additional hydraulic bearings are installed in front of each actuator. A hydraulic power station generates a flow of 330 l/min. Closed-loop control for each actuator is provided by a digital controller, Schenck 59 serial hydropuls. The controller monitors and provides signal conditioning for each load actuator position channel. Each axis (X, Y and Z) has its own dedicated strain channel for signal conditioning and control. Strain input signals can come from a variety of strain measuring devices (e.g., strain gages, extensometer). Command waveform generation for each channel can be provided either by the controller's internal waveform generators or externally via a personal computer.

Computer test control and data acquisition are performed by an object-oriented programming software (LabVIEW®). The digital controller in combination with this software package provides a highly versatile capability where numerous custom-made tests can be developed. The digital controller allows each actuator to be driven independently or in centroid control. A centroid mode uses a relationship between two opposite actuators along the same axis to maintain the center motionless. Ensuring that the center does not move prevents off-axis loading that would damage the actuators.

The digital controller enforces a centroid mode by using special algorithms to drive the servocontrol of each actuator. Figure 2 illustrates how the centroid control algorithms work for a displacement-controlled test. For each axis pair (e.g., Y+ and Y-), the controller uses the sum of the displacements and the difference of the displacements along the considered axis. This centroid control mode has been used in all the multiaxial tests presented hereafter.

## 3. Displacement field measurement by digital correlation

### 3.a Preliminaries: Correlation of Two Signals

To determine the displacement field of one image with respect to a reference image, one considers a sub-image (i.e., a square region) which will be referred to as zone of interest (ZOI). The aim of the correlation method is to match the ZOI in the two images (Fig. 3). The displacement of one ZOI with respect to the other one is a two-dimensional shift of an intensity signal digitized by a CCD camera. For the sake of simplicity, in the remainder of this section the analysis is illustrated on one-dimensional signals. To estimate a shift between two signals, one of the standard approaches utilizes a correlation function. One considers signals $g(\xi)$ which are merely perturbations of a shifted copy $f(\xi-\delta)$ of some reference signal $f(\xi)$

$$g(\xi) = f(\xi-\delta) + b(\xi) \qquad (1)$$

where $\delta$ is an unknown displacement and $b(\xi)$ a random noise. To evaluate the shift $\delta$, one may minimize the norm of the difference between $f(\xi-\Delta)$ and $g(\xi)$ with respect to $\Delta$

$$\min_{\Delta} \|g - f(.-\Delta)\|^2 \qquad (2)$$

If one chooses the usual quadratic norm $\|f\|^2 = \int_{-\infty}^{+\infty} f(\xi)d\xi$, the previous minimization problem is equivalent to maximizing the quantity $h(\Delta)$

$$h(\Delta) = (g*f)(\Delta) = \int_{-\infty}^{+\infty} g(\xi)f(\xi-\Delta)d\xi \qquad (3)$$

where $*$ denotes the cross-correlation product. Furthermore, when $b$ is a white noise, the previous estimate is optimal. The computation of a cross-correlation can be performed either in the original space or in the Fourier space, by using an FFT

$$g*f = \sqrt{N}\ \text{FFT}^{-1}\left(\text{FFT}[g]\ \overline{\text{FFT}[f]}\right) \qquad (4)$$

where the complex conjugate is overlined and N is the number of samples in the Fourier transform (to use 'fast' algorithms, it is required that $N = 2^n$, where n is an integer). The previous results can be generalized to two-dimensional situations (i.e., image matching). The use of the 'shifting' property enables one to 'move' a signal. Let us consider the shift operator $T_d$ defined by

$$[T_d f](\xi) = f(\xi - d) \quad (5)$$

where d is the shift parameter. The FFT of $T_d f$ becomes

$$FFT[T_d f] = E_d \, FFT[f] \quad (6)$$

where the modulation operator $E_d$ is defined by

$$[E_d f](\xi) = \exp(-2\pi j d \xi) f(\xi) \quad (7)$$

These two results constitute the basic tools for image correlation.

### 3.b Correlation Algorithm for Two-Dimensional Signals: CORRELI$^{GD}$

Two images are considered. The first one, referred to as 'reference image' and the second one, called 'deformed image.' One extracts the largest value p of a region of interest (ROI) of size $2^p \times 2^p$ pixel$^2$ centered in the reference image. The same ROI is considered in the deformed image. A first FFT correlation is performed to determine the average displacement $U_0$, $V_0$ of the deformed image with respect to the reference image. This displacement is expressed in an integer number of pixels and is obtained as the maximum of the cross-correlation function evaluated for each pixel of the ROI. This first prediction enables the evaluation of the maximum number of pixels that belong to the two images. The ROI in the deformed image is now centered at a point corresponding to the displaced center of the ROI in the reference image by an amount $U_0$, $V_0$.

The user usually chooses the size of the zones of interest (ZOI) by setting the value of $s < p$ so that the size is $2^s \times 2^s$ pixel$^2$. To map the whole image, the second parameter to choose is the shift $\delta x$ (= $\delta y$) between two consecutive ZOIs: $1 \leq \delta x \leq 2^s$ pixel. These two parameters define the mesh formed by the centers of each ZOI used to describe the displacement field. The following analysis is performed for each ZOI independently. It follows that parallel computations can be used in the present case. A first FFT correlation is carried out and a first value of the in-plane displacement correction $\Delta U$, $\Delta V$ is obtained. The values $\Delta U$, $\Delta V$ are again integer numbers so that the ZOI in the deformed image can be displaced by an additional amount $\Delta U$, $\Delta V$. The absolute displacement residuals are now less than 1/2 pixel in each direction. A sub-pixel iterative scheme can be used. To get good localization properties of the Fourier transform, the considered ZOI is then windowed by a modified Hanning window

$$\underline{ZOI} = ZOI \; H \otimes H \quad (8)$$

where $\underline{ZOI}$ denotes the windowed ZOI, $\otimes$ the dyadic product and H the one-dimensional modified Hanning window

$$H(i) = \begin{cases} \frac{1}{2}\left[1 - \cos\left(\frac{4\pi i}{2^s - 1}\right)\right] & \text{when} \quad 0 < i < 2^{s-2} \\ 1 & \text{when} \quad 2^{s-2} < i < 3 \times 2^{s-2} \\ \frac{1}{2}\left[1 - \cos\left(\frac{4\pi i}{2^s - 1}\right)\right] & \text{when} \quad 3 \times 2^{s-2} < i < 2^s - 1 \end{cases} \quad (9)$$

The parameter $2^{s-2}$ is an optimal value to minimize the error due to edge effects and to have a sufficiently large number of data unaltered by the window (Hild et al. 1999). A cross-correlation is performed. A sub-pixel correction of the displacement $\delta U$, $\delta V$ is obtained by determining the maximum of a parabolic interpolation of the correlation function. The interpolation is performed by considering the maximum pixel and its eight nearest neighbors. Therefore, one obtains a *sub-pixel* value. By using the 'shifting' property of the Fourier transform, one can move the deformed ZOI by an amount $\delta U$, $\delta V$. Since an interpolation was used, one may induce some errors requiring to re-iterate by considering the new 'deformed' ZOI until a convergence criterion is reached. The criterion checks whether the maximum of the interpolated correlation function increases as the number of iterations increases. Otherwise, the iteration scheme is stopped. The procedure, CORRELI$^{GD}$ (Hild et al., 1999), is implemented in Matlab$^{TM}$ (1999). The precision of the method is at least of the order of 2/100 pixel and the minimum detectable displacement is also of the order of 1/100 pixel. For a heterogeneous strain field, it is found that the precision on strain measurements is of the order of $10^{-4}$ (Hild et al., 1999).

### 3.c Finite Strain Measurements

For large displacements, two problems are to be solved: (i) the maximum displacement related to the ZOI size; (ii) the choice of the strain measure.

When the same reference picture is used, it is not possible to measure large displacements in a sequence of pictures. The solution is to follow the mesh by changing the reference view when displacement becomes too big. When large displacements are supposed to occur (i.e., in the present case) and a whole sequence of pictures is analyzed, the displacement field is determined incrementally (i.e., by considering two consecutive images of the sequence). This procedure can be simplified when the displacements remain small enough (i.e., the first image remains the reference image throughout the whole analysis).

The first point will be further discussed by using a biaxial configuration presented later on. The second question of this section is the choice of the strain tensor. It is well-known that the infinitesimal strain tensor $\underline{\underline{\varepsilon}}$ is not well adapted to large displacements for many reasons. One of them is that there is no additivity. It follows that logarithmic strains are easier to use from this point of view. Furthermore, for rigid body motion, the infinitesimal strains can be different from zero in case of pure rigid displacements. Let us examine the case of a pure rotation (see Fig. 4) by a 5° angle. In that case, the displacement field becomes:

$$\underline{u} = (\underline{\underline{R}} - \underline{\underline{1}})\underline{x} \tag{10}$$

where $\underline{\underline{R}}$ is an orthogonal second rank tensor, $\underline{x}$ the initial position with respect to the rotation axis and $\underline{\underline{1}}$ the unit second rank tensor. The infinitesimal strain tensor can be evaluated as:

$$\underline{\underline{\varepsilon}} = \frac{1}{2}(\underline{\underline{R}} + \underline{\underline{R}}^T) - \underline{\underline{1}} \neq \underline{\underline{0}} \tag{11}$$

When the rotation is measured by one angle $\theta$ in the x-y plane, the non vanishing strain components are $\varepsilon_{xx}$ and $\varepsilon_{yy}$:

$$\varepsilon_{xx} = \varepsilon_{yy} = \cos\theta - 1 \tag{12}$$

CORRELI$^{GD}$ gives an average value of both terms equal to $-3.6.10^{-3}$ when s = 6 and $\delta x$ = 32 and $-4.10^{-3}$ when s = 7 and $\delta x$ = 32. The exact value for $\theta = 5°$ is $-3.8.10^{-3}$. The shear component is equal to or less than $2.10^{-4}$. These results confirm that the precision of the strain measurements of the order of a $2.10^{-4}$ sensitivity. This incorporates the pixel interpolation algorithm used to rotate images by using PhotoShop® (1998).

Despite this good result, it also confirms the necessity to use finite strain measures when large rigid body motions are suspected to occur in addition to infinitesimal strains. These measures are built by using the gradient $\underline{\underline{F}}$ of the transformation and for the Lagrangian measures, they can be expressed as:

$$\underline{\underline{E}}_m = \begin{cases} \dfrac{1}{2m}\left(\underline{\underline{C}}^m - \underline{\underline{1}}\right) & \text{when } m \neq 0 \\ \dfrac{1}{2}\ln(\underline{\underline{C}}) & \text{when } m = 0 \end{cases} \quad \text{with } \underline{\underline{C}} = \underline{\underline{F}}^t.\underline{\underline{F}} \qquad (13)$$

where $\underline{\underline{C}}$ denotes the right Cauchy-Green tensor. When m = 1 the Green-Lagrange tensor is obtained, m=1/2 is the nominal strain tensor and yields $\Delta L/L_o$ for uniaxial elongation, m = 0 is the logarithmic strain tensor; we will use this measure when not specified. By definition, all these strains are equal to zero for a rigid rotation ($\underline{\underline{F}}=\underline{\underline{R}}$ and $\underline{\underline{C}}=\underline{\underline{1}}$). The latter property can be used to re-analyze the previous example of pure rotation. The average value of each component of Green-Lagrange or logarithmic strain tensors is less than $2.10^{-4}$. This result is in accordance with the previous sensitivity analysis in the framework of infinitesimal strains.

### *4. Study of a Rubber-Like Material*

Smactane$^{TM}$ is a rubber-like material developed by SMAC (Toulon – France). Various stiffnesses and visco-elastic performances are proposed in sheet form or as final injected product. This material was developed to obtain a maximum damping effect for a large range of temperatures (i.e., from −50° to +120°C). One finds applications in the field of flexible transmission joints, suspensions of machine, shock absorber. The grade used in this study presents virtually no damping effect and its behavior is reversible over an elongation range of 800%. An ultimate strength of 9.5 MPa (Piola-Kirchhof stress) is given by the manufacturer. During the tests, we will limit ourselves to lower stress levels to be sure that no other effect than hyper-elasticity (e.g., irreversible effects, damage) occurs.

### 4.a Hyper-Elasticity of a Rubber-Like Material

The macroscopic approach of homogeneous, hyper-elastic media such as rubber-like materials consists in the introduction of an elastic potential (Rivlin 1948). Classical hypotheses of isotropy, material-frame invariance and incompressibility allows one to assume that the state potential W only depends on the first two invariant $I_1$ and $I_2$ of the right Cauchy-Green tensor $\underline{\underline{C}}$

$$\left.\begin{aligned} I_1 &= \text{tr}\underline{\underline{C}} = \lambda_1^2 + \lambda_2^2 + \lambda_3^2 \\ I_2 &= \frac{1}{2}\left(\text{tr}^2\underline{\underline{C}} - \text{tr}\underline{\underline{C}}^2\right) = \lambda_1^2\lambda_2^2 + \lambda_2^2\lambda_3^2 + \lambda_3^2\lambda_1^2 \end{aligned}\right\} \text{ with } I_3 = \det\underline{\underline{C}} = \lambda_1\lambda_2\lambda_3 = 1 \quad (14)$$

where $\lambda_1$, $\lambda_2$, $\lambda_3$ are the three principal elongations. With such a potential, and by using the conventional formalism of continuum thermodynamics, the hyper-elastic constitutive law derives from W

$$\underline{\underline{S}} = \frac{\partial W}{\partial \underline{\underline{E}}} = 2\frac{\partial W}{\partial \underline{\underline{C}}} = 2\left(\underline{\underline{1}}\frac{\partial W}{\partial I_1} - \underline{\underline{C}}^{-2}\frac{\partial W}{\partial I_2}\right) \quad (15)$$

where $\underline{\underline{E}}$ is Green-Lagrange strain tensor and $\underline{\underline{S}}$ the second Piola-Kirchhoff extra stress tensor. The Cauchy extra stress tensor $\underline{\underline{\Sigma}}$ is related to the Piola Kirchhoff extra stress tensor by

$$\underline{\underline{\Sigma}} = \underline{\underline{F}}.\underline{\underline{S}}.{}^t\underline{\underline{F}} = 2\left(\underline{\underline{B}}\frac{\partial W}{\partial I_1} - \underline{\underline{B}}^{-1}\frac{\partial W}{\partial I_2}\right) \quad (16)$$

We obtain the complete Cauchy stress tensor $\underline{\underline{\sigma}}$ by using the relationship

$$\underline{\underline{\sigma}} = \underline{\underline{\Sigma}} - p\underline{\underline{I}} \quad (17)$$

where p is the pressure associated with the incompressibility condition. Since the partial derivatives of W with respect to $I_1$ and $I_2$ are known, the behavior is completely defined when it remains hyper-elastic.

In this section, we discuss different potential expressions presented in Table 2. We can see the forms proposed by Mooney (1940), Isihara (1951) and the neo-hookean (1941) formalism are particular cases of the general Rivlin expression. This general form is also the only one that leads to a first derivative $\partial W/\partial I_1$ that depends upon $I_1$ and $I_2$ as well as $\partial W/\partial I_2$. It is worth noting that it may not be the case if we only consider the $C_{io}$ and $C_{oj}$ terms of the expansion. For all other forms, the expressions of the partial derivatives of W with respect to $I_1$ and $I_2$ are uncoupled. As it is proposed by Lambert-Diani and Rey (1999) we call

$$\frac{\partial W}{\partial I_1} = f(I_1) \quad \text{and} \quad \frac{\partial W}{\partial I_2} = g(I_2) \tag{18}$$

In the same paper, the authors proposed a strategy to identify both functions f and g by using uniaxial and biaxial tests. Hyper-elastic models give the following Cauchy stress components in uniaxial tension ($\sigma_{UT}$) and equibiaxial tension ($\sigma_{BT}$):

$$\sigma_{UT} = 2f(I_1)\left(\lambda^2 - \frac{1}{\lambda}\right) - 2g(I_2)\left(\frac{1}{\lambda^2} - \lambda\right) \text{ with } \begin{cases} I_1 = \lambda^2 + \frac{2}{\lambda} \\ I_2 = \frac{1}{\lambda^2} + 2\lambda \end{cases}$$

$$\sigma_{BT} = 2f(I_1)\left(\lambda^2 - \frac{1}{\lambda^4}\right) - 2g(I_2)\left(\frac{1}{\lambda^2} - \lambda^4\right) \text{ with } \begin{cases} I_1 = 2\lambda^2 + \frac{1}{\lambda^4} \\ I_2 = \frac{2}{\lambda^2} + \lambda^4 \end{cases} \tag{19}$$

Figure 5 shows that $I_1$ is greater than $I_2$ in uniaxial tension and that $I_2$ is greater than $I_1$ in equibiaxial tension. We can also see that pure shear or plane strain tests lead to identical values of $I_1$ and $I_2$. By using these results some authors (Heuillet 1997 for example) propose to use such tests to identify the potential parameters. Since $I_1$ is greater than $I_2$ in uniaxial tension, an approximate identification of the function $f(I_1)$ from uniaxial data is possible and can be used to determine the function $g(I_2)$ from biaxial data (Lambert-Diani et al. 1999).

4.b Uniaxial Tension Test

A first tension test is carried out on Smactane$^{TM}$ specimen by using a classical MTS tension/compression testing machine. The tension test is carried out at room temperature (i.e., T = 22°C) and low strain rate (i.e., $\dot{\varepsilon}$ = $10^{-3}$ s$^{-1}$). A typical result of displacement field obtained by CORRELI$^{GD}$ is shown in Fig. 6 where one grip is fixed and the second one moves from the right to the left on the picture. The displacements are large but the maximum strain is only about 33%.

The problem now is that the initial mesh can move out of the picture so that for very large strains we must choose a suitable initial mesh, evaluate the displacement step by step to produce the tension curve, or move the CCD camera according to the previous displacement value. By using ASTREE and the centroid mode enforces that the center of the specimen is motionless. This point will be of extreme importance for biaxial tests (see Section 4.c).

Experimental data obtained on a Smactane[TM] specimen loaded in uniaxial tension are shown in Fig. 7. Elongation is measured by using CORRELI[GD] and the average strain of the ROI is obtained by using the CORRELI[GAUGE] procedure. The displacement field is first interpolated by using cubic B-Spline functions. The strain field components are then derived and the average values of the in-plane components are evaluated. The pictures of Fig. 6 are related to the first four points of Fig. 7 from which we can evaluate the tangent Young's modulus E. This modulus is equal to 3.3 MPa.

If we assume that the contribution of the second invariant is negligible with respect to the first one in a uniaxial stress $\sigma_{UT}$ (since $I_1 \gg I_2$), the function f can be directly identified from the knowledge of the Cauchy stress and elongation $\lambda$

$$f(I_1) = \frac{\sigma_{UT}}{2\left(\lambda^2 - \frac{1}{\lambda}\right)} \quad \text{with} \quad I_1 = \lambda^2 + \frac{2}{\lambda} \tag{20}$$

Figure 8 shows that this function is not constant during the tensile test on Smactane[TM]. Apart from the models proposed by Hart-Smith (1966) and Alexander (1968), all the others will not be able to fit the experimental data. The two models lead to similar expressions of the first invariant function

$$\ln[f(I_1)] = \ln G + k_1(I_1 - 3)^2 \tag{21}$$

where G and $k_1$ are material parameters. To fit the experimental data obtained by Treolar (1944) data, Lambert-Diani and Rey (1999) generalized this form

$$f(I_1) = \exp\left\{\sum_{i=0}^{n} a_i (I_1 - 3)^i\right\} \tag{22}$$

where $a_i$ are material parameters. These forms are compared in Fig. 8 for Smactane[TM]. We choose the value n = 2 because the experimental data are well described and the number of parameters is minimal, even though the value n = 3 fits better the data. It is worth noting that the form proposed by Rivlin assuming uncoupled evolution of f and g, would give as accurate representation of the experimental data when choosing the same number of parameters. The different coefficients are summarized in Table 3.

The next question to address is the determination of the second invariant function $g(I_2)$. By using a conventional uniaxial testing machine, a commonly used procedure is to design large specimen (initial width $L_o$) with respect to the initial height $h_o$ (see Fig. 9). Since the ratio h/L remains "small," a large area of the specimen is in a plane strain state (i.e., $\varepsilon_2 = 0$). Edge effects ($\sigma_2 = 0$) induce perturbations that propagate inside the specimen over a length $\delta$ on both sides. Numerical simulations as well as video control on experimental "plane strain" tests prove that $\delta$ is of the order of the specimen length h. In this zone the axial Cauchy stress decreases from $\sigma_1$ to $\kappa\sigma_1$ where $\kappa$ value is about 0.6 for small strain levels.

If $\kappa$ remains constant during the test, the error on stress measurement increases with extension ratio $\lambda$ and quickly reaches important values (i.e., up to 100%). The parameter $\kappa$ increases from 0.6 value to 1 when $\lambda$ reaches about 5. The stress error decreases to zero (i.e., the stress distribution becomes homogeneous in the specimen) but we no longer are in plane strain tension. The strain field is similar to a simple tension strain field. Besides, the invariants $I_1$ and $I_2$ are equal during pure plane strain tension so that the effect of the function g is not dominant for such a test. Consequently, a biaxial testing procedure is developed to identify the second part of the hyper-elastic potential.

4.c    Biaxial Tension Test

The material is equally stretched in both directions of the sheet (Fig.10). Fig. 10-b shows typical CORRELI$^{GD}$ results obtained when a bad positioning of the specimen in the grips occurs. The effect of rotation and elongation is superimposed with edge effects. This picture is typical of asymmetric boundary conditions. CORRELI$^{GD}$ is of great help to validate the positioning of the specimen. The specimen is slightly stretched and an analysis is run. It takes at most a few minutes to get a full displacement field and to check its homogeneity. It follows that the experimental boundary conditions can be checked a priori and used a posteriori to simulate the experiment.

It is easy to understand that the first evaluation of the displacement vector obtained by correlation of the ZOI will not be accurate as soon as the effective displacement is greater than half the length of the ZOI. In Fig. 11a we can observe that the evaluation of the displacement field is not satisfactory. The biaxial extension of the plane specimen is about 7% in both directions. With a 32 pixel ZOI (i.e., s = 5), only a small square region leads to reasonable results of the correlation technique in agreement with the previous remark (the range of the displacement contours is equal to 40 pixel). This zone increases with a 64 pixel ZOI (the displacement range is about 2/3 of the size of the ZOI). Some artifacts remain in the lower part of Fig. 11b. The solution is to choose a larger ZOI (s = 7) when the displacement becomes important. Figure 11c shows that the problem is solved and that the displacement field is linear in both directions. Table 1 shows the average strains over the whole ROI for the three sizes of ZOI. When s = 5, the average longitudinal strains are negative even though the specimen was stretched. This first result clearly shows again that s = 5 is too small for this displacement and strain range. The value s = 6 leads to reasonable results. However, one may note that the strains are different in the two directions. When s = 7 the shear strain is less than precision of the correlation technique. The longitudinal strains are virtually identical in the two directions: this experiment is indeed representative of an equibiaxial stretching. It is worth noting the very good homogeneity of the strain field with such a simple set-up. We will see that this homogeneity is also observed even with larger displacements.

Figures 12a, 12b and 12c show the displacement field after biaxial testing on a Smactane$^{TM}$ specimen. The corresponding strain field is not completely homogeneous (Fig. 12d). By considering the strain quasi-constant along AB (see Fig. 12a and 12c), we assume that the stresses are uniform in this section AB. Consequently, we can estimate the Cauchy stress (i.e., $\sigma_{BT} = \sqrt{2}$ F/eL where e is the current specimen thickness and L the current length between A and B). The result is plotted in Fig. 13 from which we can deduce the initial slope. This slope value is equal to 7.6 MPa that is less than 10% different from the theoretical value '2E' for uncompressible material.

The biaxial Cauchy stress helps us to identify the form of the second invariant function $g(I_2)$ by using Eqn. (23)

$$g(I_2) = \frac{\sigma_{TB} - 2f(I_1)\left(\lambda^2 - \frac{1}{\lambda^4}\right)}{2\left(\lambda^4 - \frac{1}{\lambda^2}\right)} \quad \text{with } I_1 = 2\lambda^2 + \frac{1}{\lambda^4} \text{ and } I_2 = \lambda^4 + \frac{2}{\lambda^2} \quad (23)$$

Figure 14 shows the shape of the function $g(I_2)$ which can be fitted by the models proposed by Gent and Thomas (1958), Hart-Smith (1966) or Alexander (1968). Two different forms are analyzed in this figure: (i) the Hart-Smith form; (ii) the Lambert-Diani and Rey form, respectively

$$g(I_2) = \sum_{i=0}^{n} \frac{c_i}{I_2^{i}} \tag{24}$$

$$g(I_2) = \exp\left(\sum_{i=0}^{n} b_i (I_2 - 3)^i\right) \tag{25}$$

But the identification does not stop here; CORRELI$^{GD}$ evaluates the whole strain field at each step of the test and by using the estimated material coefficients we can calculate the stress field. It follows that the forces in each direction can be computed and compared with measured load. This method leads to optimize material parameters for the chosen constitutive law. By inverse analysis of the problem, we modify these coefficients to fit as well as possible the experimental data. In Fig. 15, the simulations using the Lambert-Diani and Rey model are compared with the uniaxial data. In the same figure, the simulation using only the f function contribution is plotted. A posteriori, one can observe the small influence of the second invariant function g, on the uniaxial simulation since experimental data are well fitted with the complete model (i.e., f and g contributions.)

## 5. Conclusions and Perspectives

A multiaxial testing procedure has been used to analyze the behavior of a rubber-like material. By using a CCD camera coupled with digital image correlation, we can measure the displacement field over all the specimen. This approach is particularly well-adapted for large displacement tests where it is difficult to obtain homogeneous strain fields. Such a non intrusive method helps one to analyze more thoroughly experiments on rubber-like as well as other materials. In particular, the experimental boundary conditions can be checked.

Uniaxial and equibiaxial tests have been carried out on Smactane$^{TM}$. The results are analyzed by assuming that the contribution of the invariants $I_1$ and $I_2$ are uncoupled in the hyper-elastic potential. Experimental data are used to discriminate classical hyper-elastic models. For a Smactane$^{TM}$ material, the Mooney-Rivlin model is not representative of the hyper-elastic behavior when strains over 100% occur. The models proposed by Alexander or Hart-Smith are more representative even though the time-dependent part of the behavior was not studied. By using such models, we can simulate other loading conditions (e.g., plane strain, biaxial tension) and compare deformed shapes and strain fields with numerical simulations of the test. This constitutes one of the perspectives to this work. Another perspective is to take into account the strain rate effects (by comparing successive strain field measurements for small time intervals). Such an analysis can be used to analyze the visco-elastic properties of rubber-like materials and other polymers.

## *Acknowledgments*

The authors wish to thank Dr. Jean-Noël Périé who has been of great help in the CORRELI programming under the Matlab$^{5.3}$ environment. SMAC-Toulon provided us with Smactane$^{TM}$ sheets.

*List of tables*

*List of figures*

Figure 10: Positioning of the specimen during biaxial testing in the multiaxial machine ASTREE.

Figure 11: Influence of ZOI size on the evaluation of the displacement field in a biaxial test.

Figure 12: Biaxial stress measurement, displacement and strain field in a biaxial test on Smactane$^{TM}$.

Figure 13: Stress/strain response of Smactane$^{TM}$ in uniaxial and equibiaxial tension.

Figure 14: Second invariant function identification.

Figure 15: Description of the hyper-elastic behavior of Smactane$^{TM}$.

|       | $\varepsilon_{xx}$ (%) | $\varepsilon_{yy}$ (%) | $\varepsilon_{xy}$ (%) |
|:-----:|:----------------------:|:----------------------:|:----------------------:|
| s = 5 | -0.10                  | -0.40                  | 0.09                   |
| s = 6 | 5.50                   | 4.50                   | 0.37                   |
| s = 7 | 5.47                   | 5.42                   | <0.01                  |

Table 1: Average in-plane strains for three different sizes of ZOI.

| Author(s) | Hyper-elastic potential | Number of material parameters |
|---|---|---|
| Mooney (1940) | $W = C_1(I_1-3) + C_2(I_2-3)$ | 2 |
| Neo-Hookeen (1943) | $W = G/2\,(I_1-3)$ | 1 |
| Rivlin (1943) | $W = C_{ij}(I_1-3)^i(I_2-3)^j$ | ? |
| Isihara (1951) | $W = C_1(I_1-3) + C_2(I_2-3)^2 + C_3(I_2-3)$ | 3 |
| Rivlin, Saunders (1951) | $W = C_1(I_1-3) + f(I_2-3)$   with: $f(0)=0$ | ? |
| Gent and Thomas (1958) | $W = C_1(I_1-3) + C_2 \ln(I_2/3)$ | 2 |
| Biderman (1958) | $W = C_{10}(I_1-3) + C_{20}(I_1-3)^2 + C_{30}(I_1-3)^3 + C_{01}(I_2-3)$ | 4 |
| Hart-Smith (1966) | W as: $\dfrac{\partial W}{\partial I_1} = G \exp\left(k_1(I_1-3)^2\right)$ and $\dfrac{\partial W}{\partial I_2} = G\dfrac{k_2}{I_2}$ | 3 |
| Alexander (1968) | $W = \dfrac{\mu}{2}\left\{C_1 \int \exp\left(k(I_1-3)^2\right)dI_1 + C_2 \ln\left(\dfrac{I_2-3+\gamma}{\gamma}\right) + C_3(I_2-3)\right\}$ | 5 |

Table 2: Hyper-elastic potential expressions.

| n = 1 | n = 2 | n = 3 | n = 4 |
|---|---|---|---|
| $a_0 = -1.38$ | $a_0 = -1.27$ | $a_0 = -1.86$ | $a_0 = -1.14$ |
| $a_1 = -1.03 \ 10^{-2}$ | $a_1 = -2.52 \ 10^{-2}$ | $a_1 = -8.01 \ 10^{-2}$ | $a_1 = -1.31 \ 10^{-1}$ |
| | $a_2 = 1.40 \ 10^{-3}$ | $a_2 = 7.10 \ 10^{-3}$ | $a_2 = 1.65 \ 10^{-2}$ |
| | | $a_3 = -1.45 \ 10^{-4}$ | $a_3 = -7.00 \ 10^{-4}$ |
| | | | $a_4 = 1.02 \ 10^{-5}$ |

Table 3: Material parameters for the function f when n = 1, 2 3 & 4 (Smactane[TM]).

| (i) n = 1 | (i) n = 2 | (ii) n = 1 | (ii) n = 2 | (ii) n = 3 |
|---|---|---|---|---|
| $c_0 = 3.25 \ 10^{-1}$ | $c_0 = -9.53 \ 10^{-3}$ | $b_0 = -1.00$ | $b_0 = -1.02$ | $b_0 = -1.03$ |
| $c_1 = 2.76 \ 10^{-1}$ | $c_1 = 2.89$ | $b_1 = -0.01$ | $b_1 = -3.92 \ 10^{-3}$ | $b_1 = -7.22 \ 10^{-3}$ |
| | $c_2 = -4.39$ | | $b_2 = -5.30 \ 10^{-4}$ | $b_2 = -2.05 \ 10^{-3}$ |
| | | | | $b_3 = -1.70 \ 10^{-4}$ |

Table 4: Material parameters for the function g (Smactane[TM]).

    (i)     Hart-Smith model when n = 1 & 2

    (ii)    Lambert-Diani and Rey model when n = 1, 2 & 3

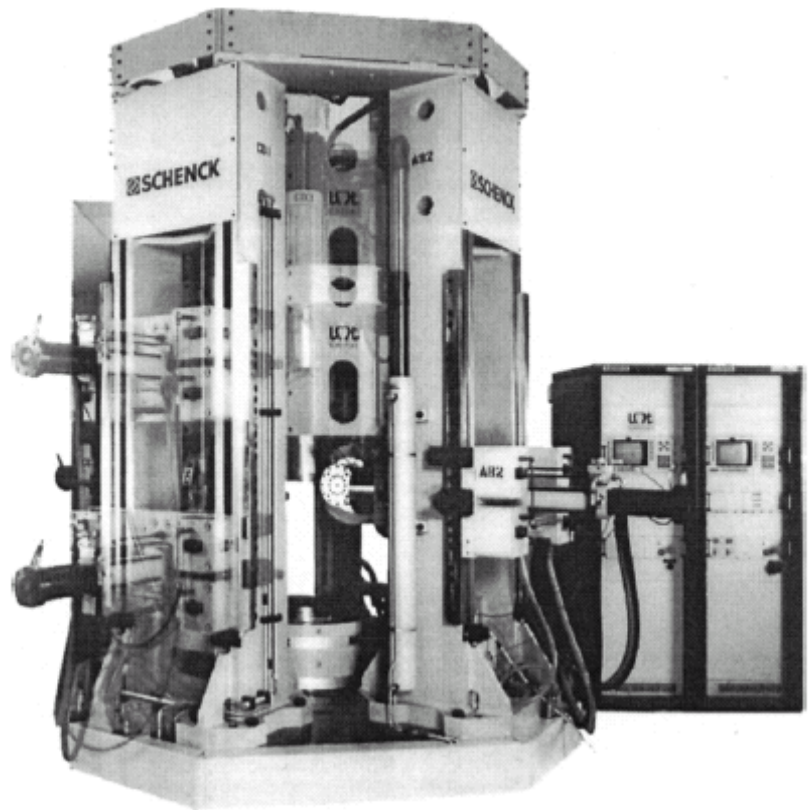

Figure 1: Multiaxial testing machine ASTREE.

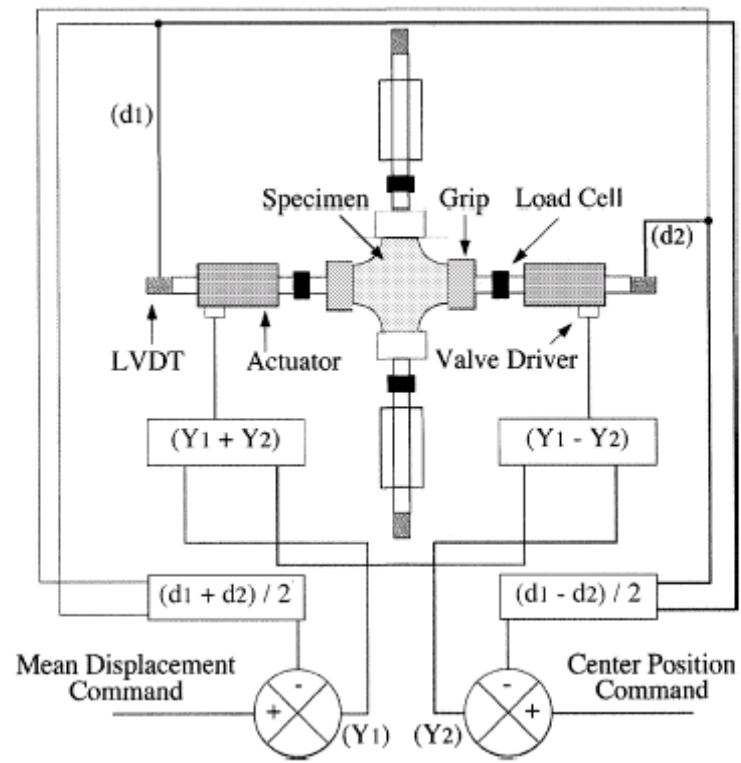

Figure 2: Coupled actuators for the multiaxial testing machine ASTREE.

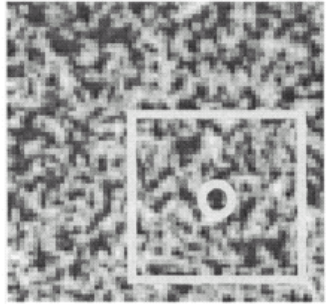 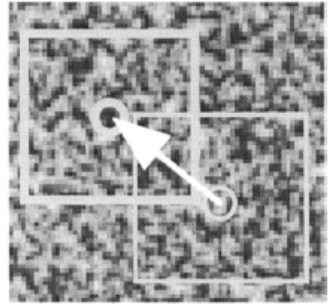

Figure 3: ZOI in the initial and 'deformed' image.

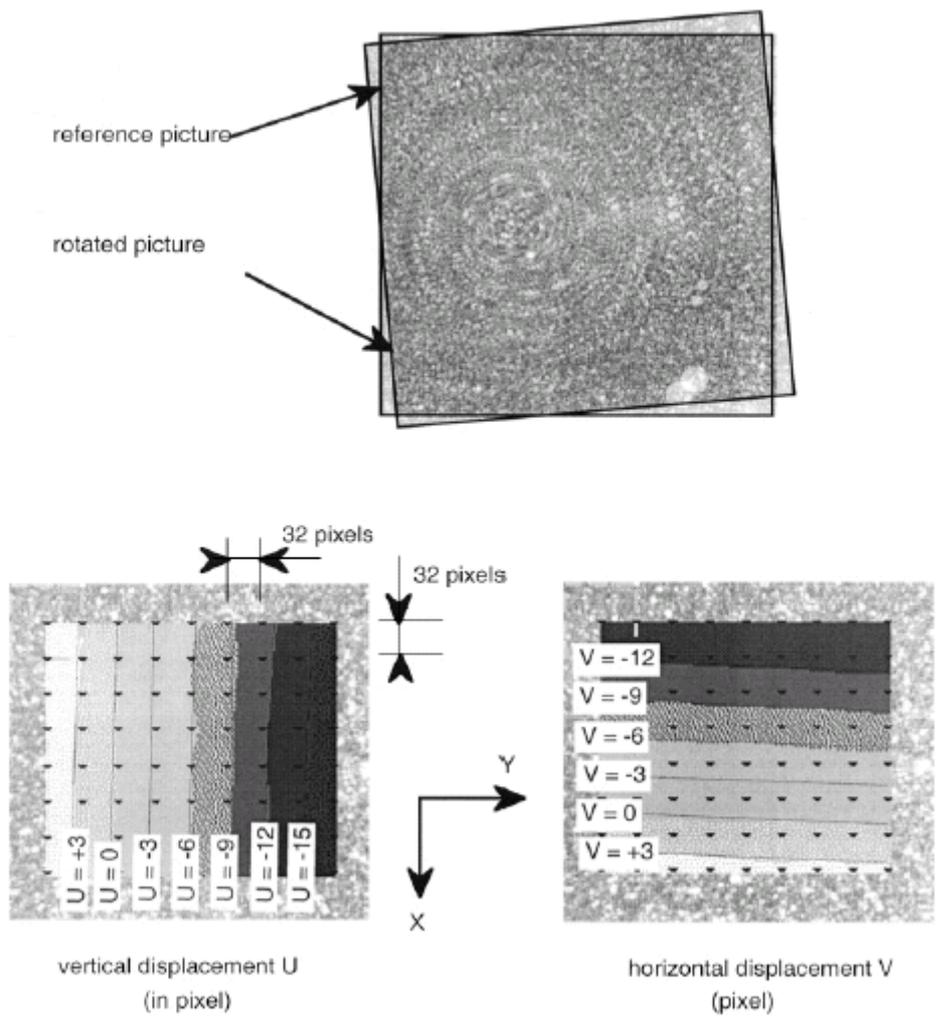

Figure 4: Displacement contours in pure rotation (s = 6 and δx = 32 pixels).

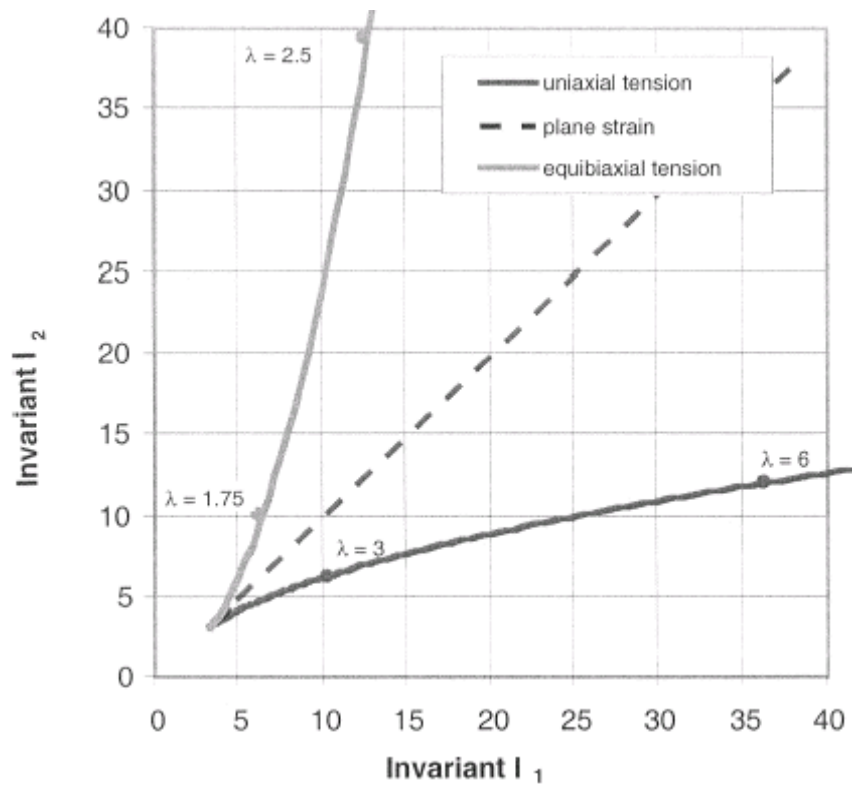

Figure 5: Invariants $I_1$ and $I_2$ for different tests.

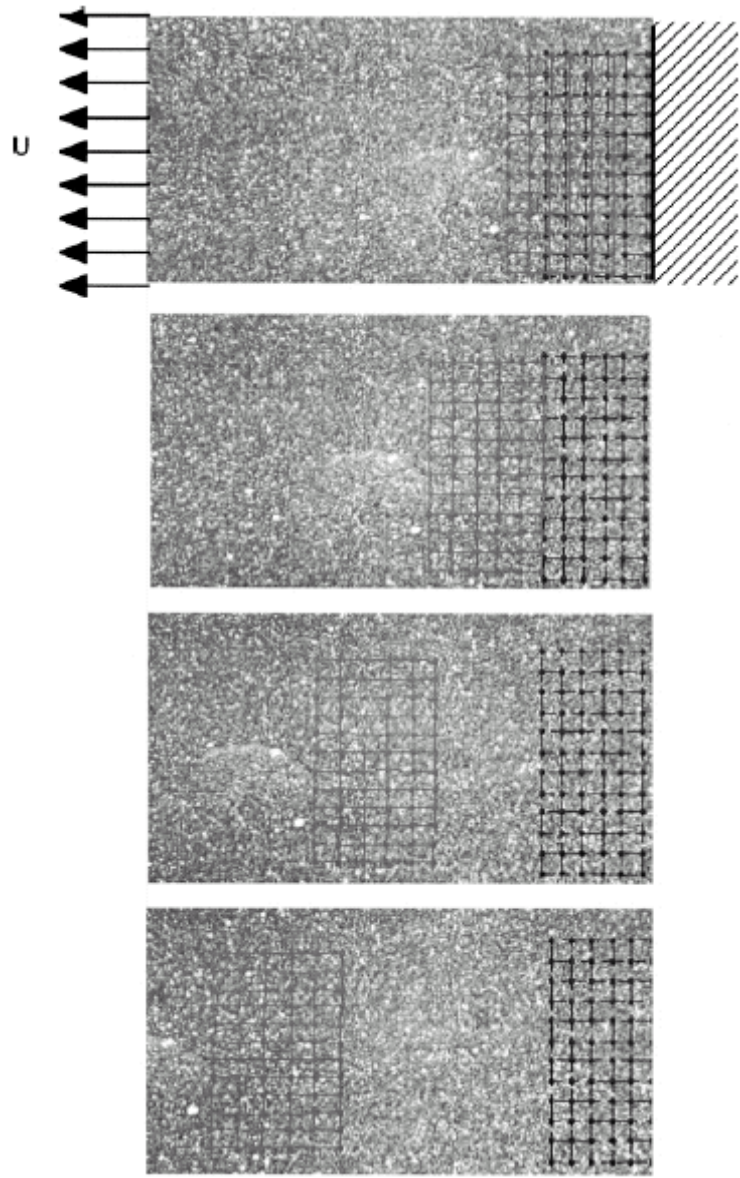

Figure 6: Large displacements measured with CORRELI$^{GD}$ in a uniaxial tensile test on Smactane$^{TM}$.

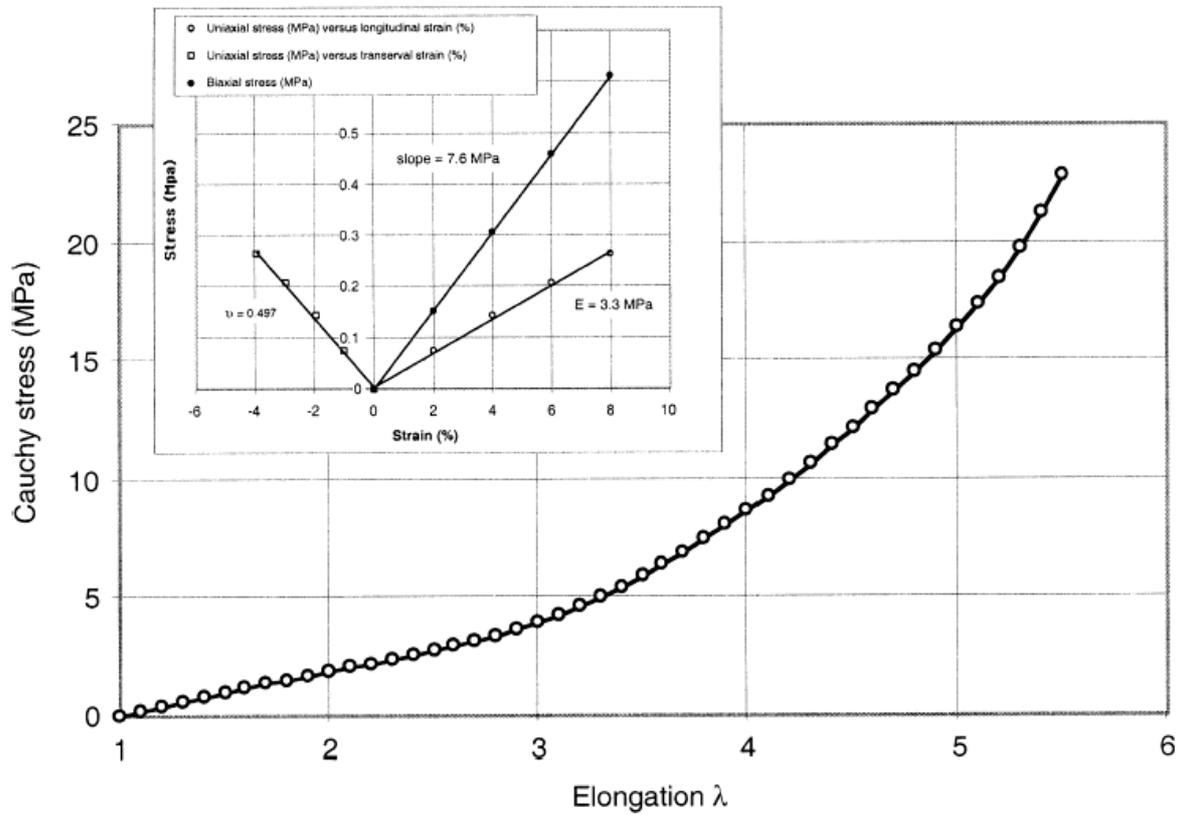

Figure 7: Stress/strain response of Smactane[TM] in uniaxial tension. Initial stress/strain curve in axial and equibiaxial tension.

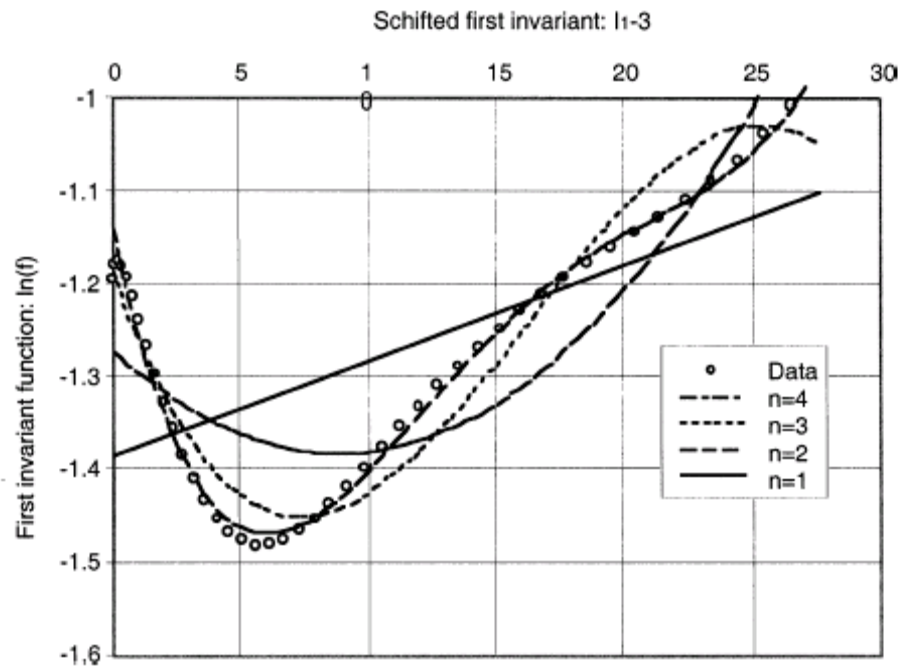

Figure 8: Identification of the first invariant function for different values of n.

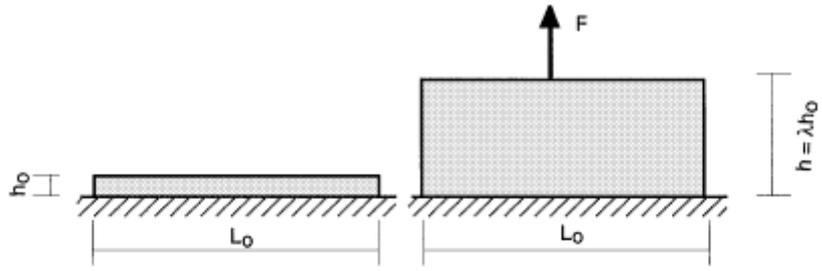

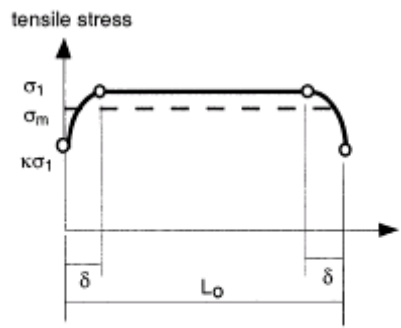

Figure 9: 'Plane strain' test, principle and limitation.

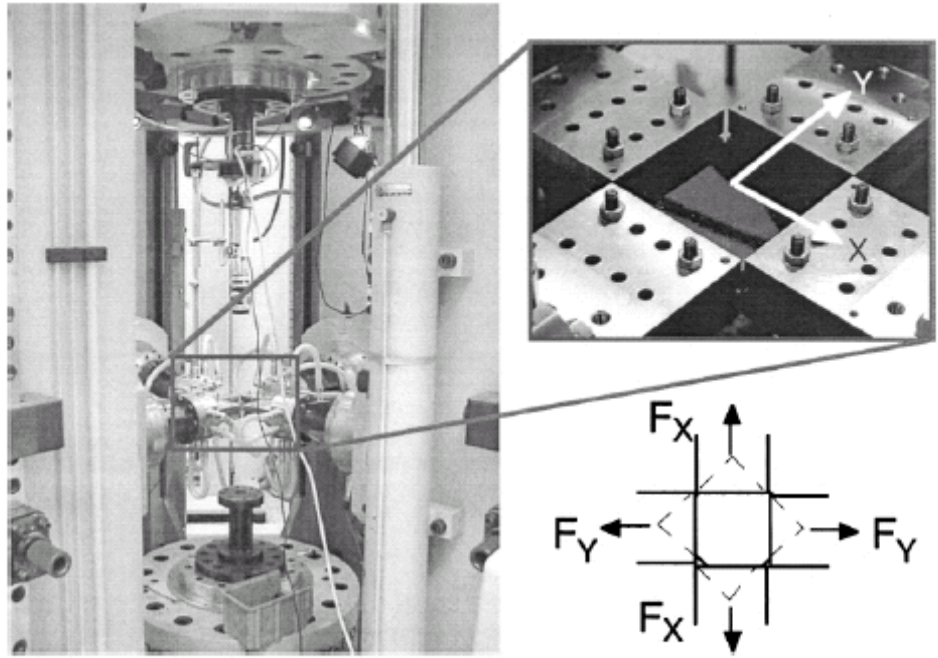

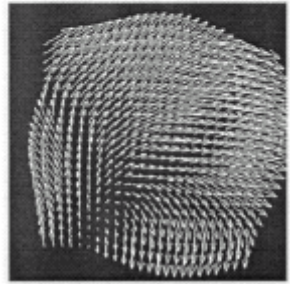

(b) Bad positioning of the specimen: displacement field (X 8)

Figure 10: Positioning of the specimen during biaxial testing in the multiaxial machine ASTREE.

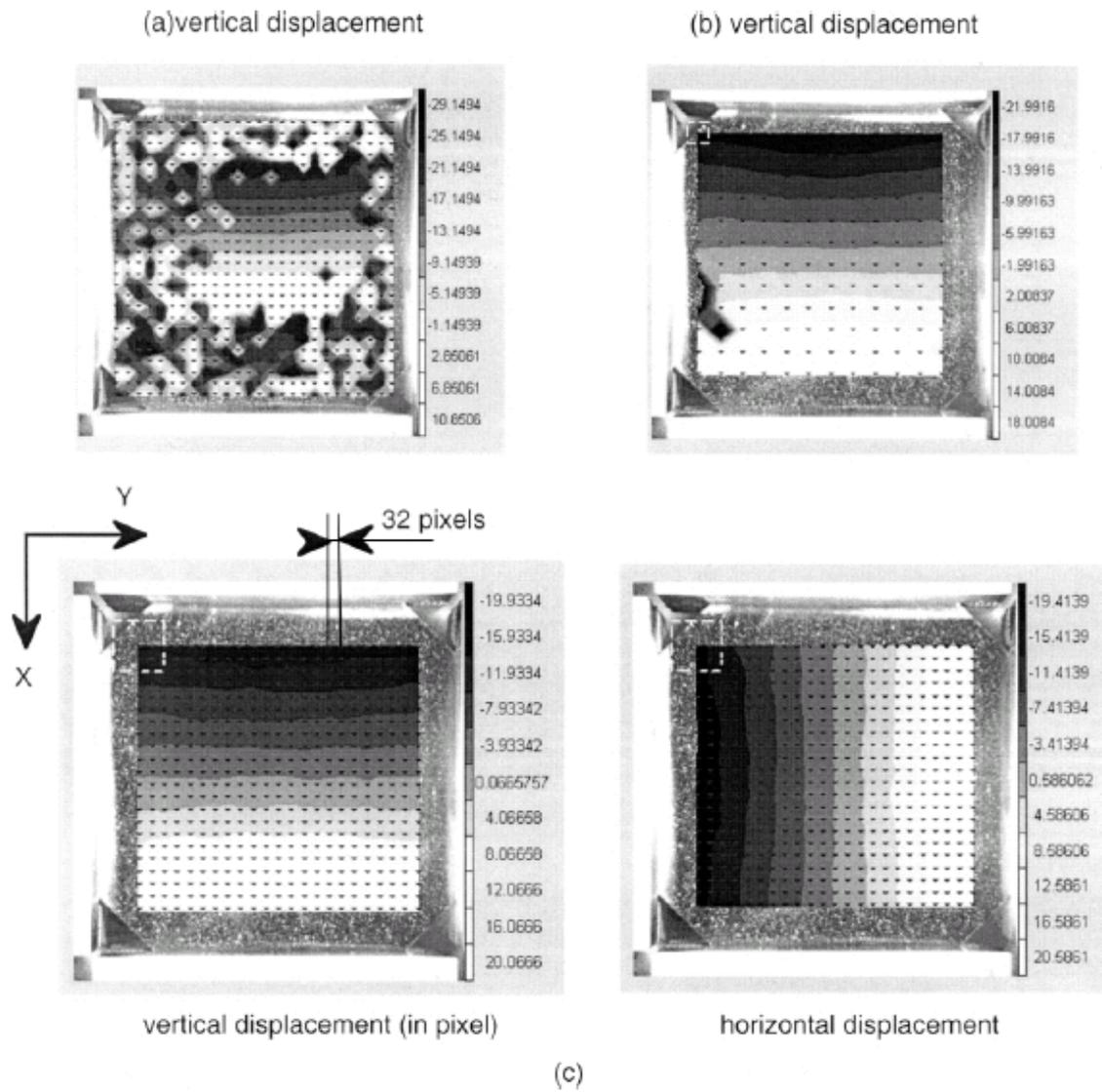

Figure 11: Influence of ZOI size on the evaluation of the displacement field in a biaxial test.

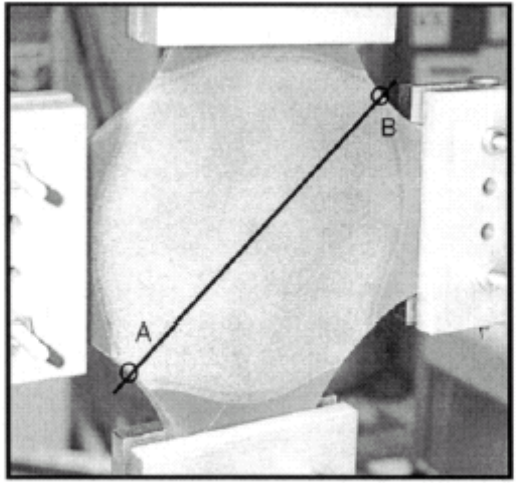
(a) Deformed specimen

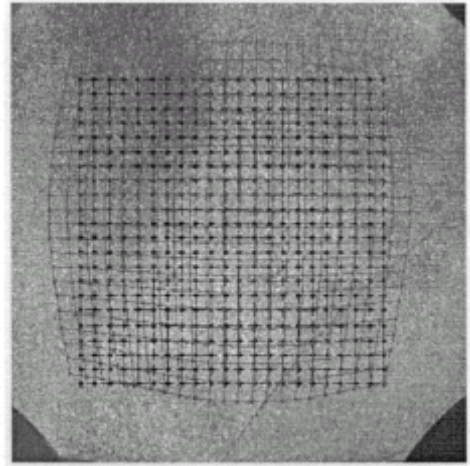
(b) Initial and deformed mesh

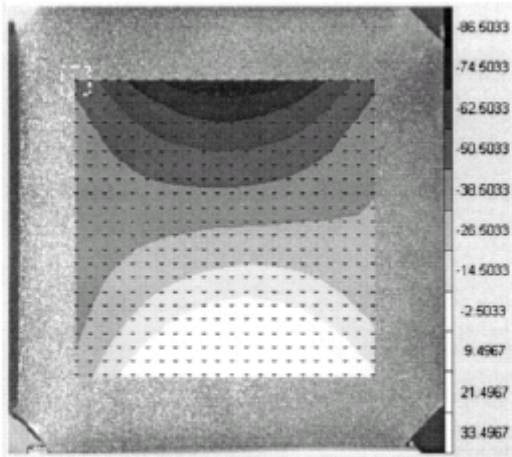
(c) Vertical displacement contours

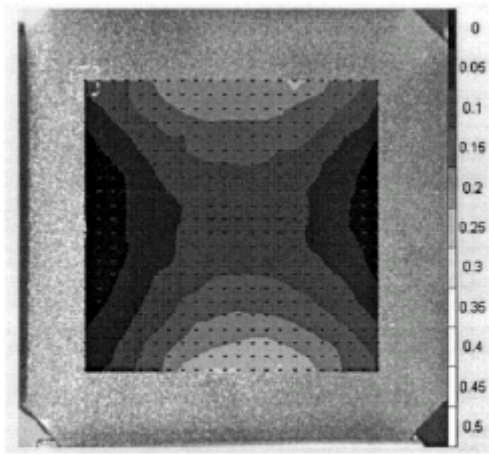
(d) Vertical Green-Lagrange strain contours

Figure 12: Biaxial stress measurement, displacement and strain field in a biaxial test on Smactane$^{TM}$.

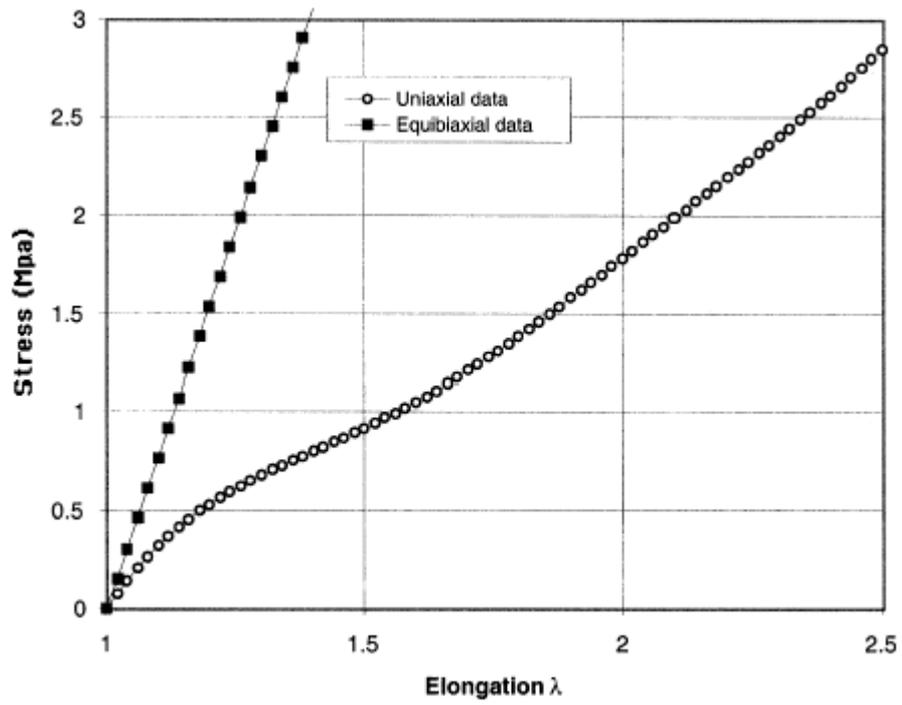

Figure 13: Stress/strain response of Smactane[TM] in uniaxial and equibiaxial tension.

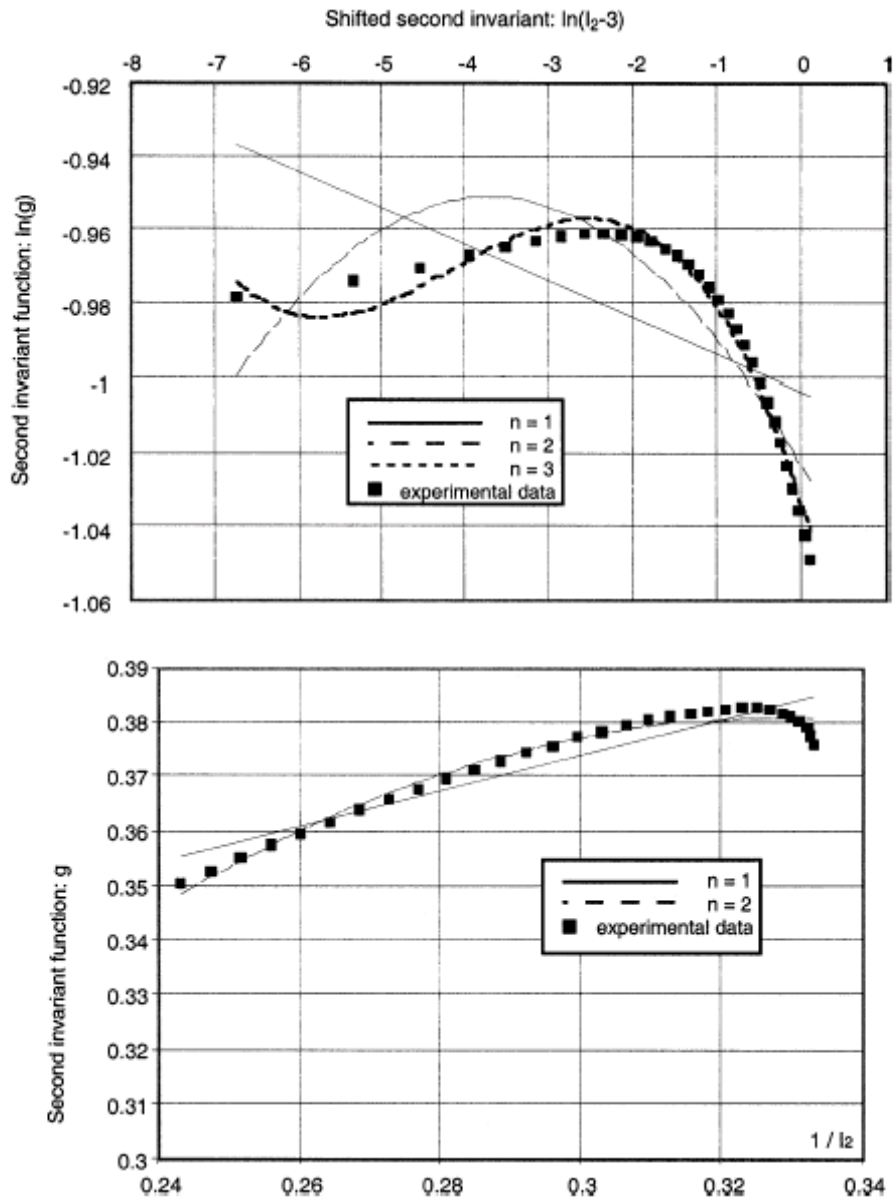

Figure 14: Second invariant function identification.

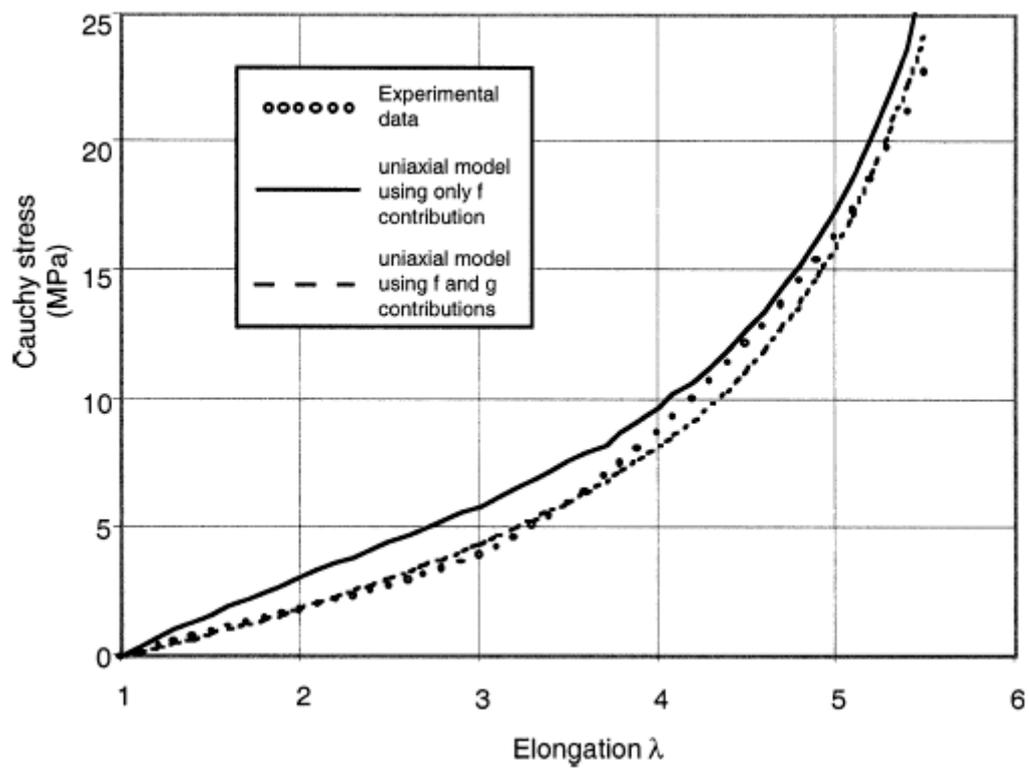

Figure 15: Description of the hyper-elastic behavior of Smactane$^{TM}$.